# Computer Simulation of Biological Processes at the High School

Olena V. Komarova[0000-0002-3476-3351]

Kryvyi Rih State Pedagogical University, 54, Gagarina Ave., Kryvyi Rih, 50086, Ukraine
`komarova1978@ukr.net`

**Abstract.** *Research goals*: the necessity of study in high school of the law of Hardy – Weinberg as one of the fundamental genetic laws was justified. The peculiarities of using the method of model experiment in the study of the genetic and evolutionary processes in populations with the use of computer technology. *Object of research*: computer simulation of population genetic structure. *Subject of research*: computer simulation of genetic and evolutionary processes in ideal and real populations. *Research methods:* pedagogical experiment (survey), analysis of scientific publications on the use of the high school method of modelling genetic and evolutionary processes in populations, computer simulation. *Results of the research:* a web page for processing by the pupils of the modelling results of genetic and evolutionary processes in populations was created.

## 1 Introduction

### 1.1 The Problem Statement

Modern course of biology in high school is based on the fundamental theoretical generalizations of basic biological science – scientific theories and laws. Fundamental genetic laws, classically studied by high school students, are laws of heredity of Mendel. Given the trends of development of modern biological sciences, namely, the development of theoretical biology, the main issues which are problems of genetics, ecology, evolution, law of genetic equilibrium concentrations (the law of Hardy – Weinberg) is considered as a fundamental law, the disclosure of which to high school students is aimed at understanding by them of the mechanism of evolution in general. This law reveals the regularities of functioning of living at the population – species level, including time frames.

## 1.2 The State of the Art

Students' mastering of the patterns of population genetics and associated evolutionary theory is one of the most complex issues in biology course in high school. Studies of such scientists as Robert L. Hammersmith, Thomas R. Mertens [1; 2; 3], Timothy J. Maret, Steven W. Rissing [4], Carol Chapnick Mukhopadhyay, Rosemary Henze, Yolanda T. Moses [5], Pongprapan Pongsophon, Vantipa Roadrangka, Alison Campbell [6] confirm this.

We have conducted a survey among 52 high school students to ascertain their level of knowledge about the essence of law of genetic equilibrium concentrations, its value for the understanding of the factors and directions of the evolutionary process.

The tasks were as following:

1. Specify the mathematical equation of the law of Hardy – Weinberg (multiple answers are allowed):

   A) $p + q = 1$;
   B) $(p + q)^2 = 1$;
   C) $p^2 + 2pq + q^2 = 1$;
   D) $p^2 + pq + q^2 = 1$;
   E) $p + 2pq + q = 1$.

2. Specify an equation describing the genotypic structure of the population (multiple answers are allowed): (see the answers to the assignment 1).
3. Specify the equation describing allelic population structure: (see answers to the assignment 1).
4. What are the conditions of validity of the law of equilibrium gene concentrations?

Students were asked to solve three problems for the application of the law of equilibrium of gene concentrations.

The results of the survey are presented in Fig. 1–3.

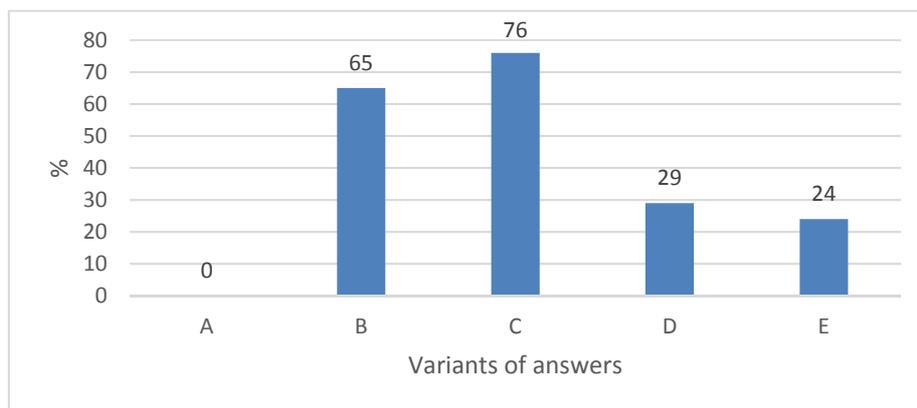

**Fig. 1.** The results of the response to the first task

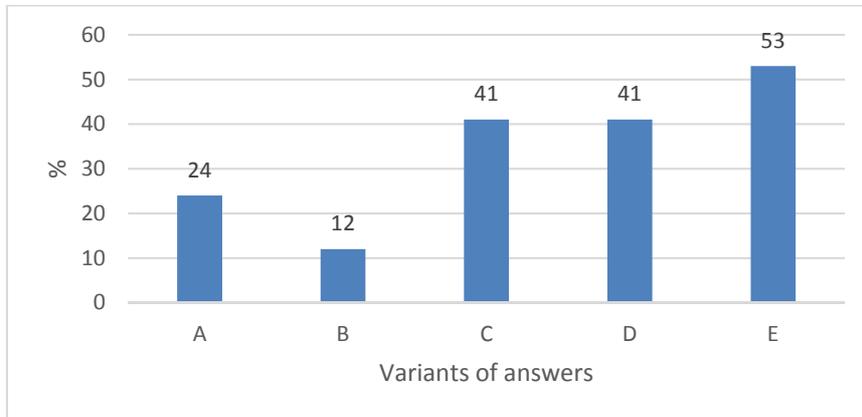

**Fig. 2.** The results of the response to the second task

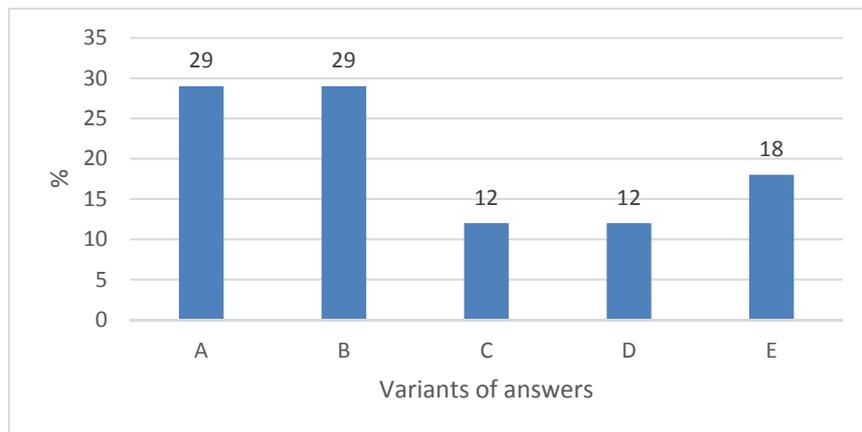

**Fig. 3.** The results of the response to the third task

Conditions of validity of the law, according to the student, were as following: population sizes are large – 29% of respondents, mating occurs at random – 24%, new mutations do not occur – 18%, all genotypes are equally fertile – 12%, generations do not overlap – 12%, there is no exchange of genes with other populations – 18%, the genes are in the autosomes and not in sex chromosomes – 18%, individuals of different genotypes are equally viable – 12%.

The obtained results allow to formulate the following conclusions: students insufficiently understood the description of the essence of the law of Hardy – Weinberg with two equations, namely, the definition of allelic and genotype structure of the population; the students are confused about variables included in the equations; knowledge about the conditions of the law is fragmentary. None of the respondents began to address two of the proposed problems, three of the respondents solved the third task incorrectly.

The results of the survey suggest the presence of formalism in high school students' knowledge about the law of genetic equilibrium concentrations. Formal approach to training lies in the mechanical memorization of educational material without enough understanding of its content. The low level of development of knowledge about the law of genetic equilibrium concentrations is one of the reasons for the difficulties of the students in understanding of the evolutionary content for the understanding of population genetics and population human genetics in particular.

Simulation, particularly computer simulation is one of the most effective training methods for demonstrating to students of the essence of complex biological processes, including genetic and evolutionary processes in natural populations.

### 1.3 The Purpose of the Article

Our main goal is to create a web pages for online processing of the modelling results of genetic and evolutionary processes in populations.

## 2 Presenting the Main Material

### 2.1 The Technological Aspect of Use

Model experiment is a special form of the experiment, which is characterized by the use of existing models as special means of experimental research. The purpose of using of the method of model experiment in the study of the genetic and evolutionary processes in populations: students get convinced in practice that, in ideal populations gene frequency and the ratio of genotypes from generation to generation are preserved, in contrast to populations influenced by the genetic factors of dynamics; simulation experiments allow us to represent the primary evolutionary changes in populations; model experiments demonstrate the probabilistic nature of genetic and microevolutionary processes; model experiments contribute to the transformation of empirical knowledge of students in persistent beliefs that are an integral component of outlook.

We have developed a web page for entering, processing and presenting graphical view of modelling results of genetic and evolutionary processes in ideal populations, which are not influenced by the factors of changing its genetic structure (according to the law of Hardy – Weinberg) – http://mybio.education/mod/exp1/en/index.html (Model experiment 1. Study of the genetic structure of the ideal population) and http://mybio.education/mod/exp2/en/index.html (Model experiment 1. Study of the genetic structure of the ideal population (second option)), as well as web pages to make for entering the results of modelling of genetic and evolutionary processes in populations, which are influenced by the factors of changing its genetic structure http://mybio.education/mod/exp3/en/index.html (Model experiment 2. Study of the genetic structure of the population under the influence of natural selection), http://mybio.education/mod/exp4/en/index.html (Model experiment 3. Modelling the effect of gene flow on the genetic structure of the population), http://mybio.education/mod/exp5/en/index.html (Model experiment 4. Modelling the effect of random processes on the genetic structure of the population, modelling the drift of genes).

The developed system of online processing of simulation results can only be used if in a model experiment the number of model individuals of the population is insignificant. The population size is limited by the objective possibility of creating a corresponding number of chip patterns of the alleles of a gene. Optimum number of chips – 100. In this case, the number of individuals is equal to 50. One can take more or fewer objects. In the first case, the choice will be associated with the growth of material costs for the manufacture of model elements. In the second case, the calculated values (allele frequency) will be significantly deviate from the pre-selected frequencies, and the level of statistical significance of the obtained results will decrease.

## 2.2 The Ways of Implementation

Stages of modelling of the genetic structure of populations are as following:

1. Modelling of the genetic structure of an ideal population with the use of material objects. Entry of simulation results into a table on web pages http://mybio.education/mod/exp1/en/index.html or http://mybio.education/mod/exp2/en/index.html.

Modelling of the genetic structure of an ideal population can be done using the possibilities of any of the two web pages. The difference between them lies in the methods of processing of the experimental results, namely in the methods of calculating the frequencies of genes. In the first variant, the gene frequencies are calculated automatically by the method of extracting of the square roots of the frequencies of the homozygotes AA and AA. In the second variant the gene frequencies are automatically calculated according to the formulas: $p = (D + 0.5 H) / N$, $q = (R + 0.5 H) / N$, where $p$ – frequency of dominant allele, $q$ – frequency of recessive allele, $D$ – number of dominant homozygotes, $R$ – number of recessive homozygotes, $H$ – number of heterozygotes, $N$ – total number of members of the population. Both methods allow us to formulate the main conclusion, that in ideal populations, the ratio of frequencies of genes and genotypes remain constant from generation to generation, and the sum of their frequencies is equal to 1.

2. Modelling of population genetic structure, which is influenced by factors of change in its genetic structure – natural selection, gene flow, genetic drift. Entry of simulation results into a table on web pages http://mybio.education/mod/exp3/en/index.html, http://mybio.education/mod/exp4/en/index.html, http://mybio.education/mod/exp5/en/index.html respectively.

Before usage of web pages for entering the results of the simulation, high school students work with persisted models of alleles of dealing a gene and create a genetic model of the parent population [2; 3]. These materialized models can be checkers, chips, candies, balls of different colours. The educational models of the genetic structure of the population are the findings of the experimental action with the model elements first ratio of genotypes and ratio of frequencies of genes, that is, the ratio of frequencies of genotypes and genes in the parent population.

On each of the web pages there is an instruction for the sequence of actions that must be performed concerning materialized objects, as well as actions to enter the received results in the tables for automatic calculation of genotype frequencies and allele frequencies. The rows that are highlighted in blue in tables for web pages http://mybio.education/mod/exp1/en/index.html, http://mybio.education/mod/exp2/en/index.html, http://mybio.education/mod/exp3/en/index.html or http://mybio.education/mod/exp4/en/index.html, http://mybio.education/mod/exp5/en/index.html are filled manually by students on the basis of counting of the number of the results obtained in the course of the materialized models of alleles and genotypes. The web pages provide automatic plotting of graphs and charts, allowing, first, to reveal the results in graphical form (Fig. 4, 5). Secondly, it allows to effectively carry out their comparative analysis and to formulate conclusions according to the algorithm of the action plan.

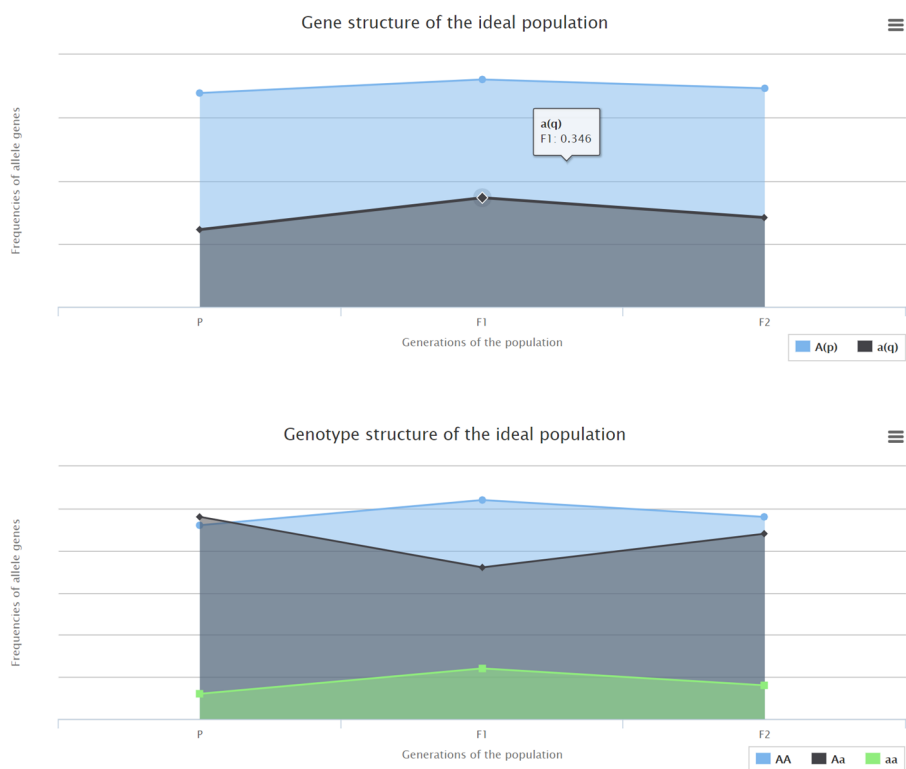

**Fig. 4.** View of graphics on the web page that were built in automatic mode http://mybio.education/mod/exp1/en/index.html#

Both diagrams show the genetic structure of populations and according to the semantic content they are identical. They differ in the way of the visibility of the results. The teacher can draw the students' attention to one variant of a diagram with a proposal to compare the genetic, genotypic structure of the population in generations. There is another, more complicated version of the analysis of the constructed diagrams. For this the students choose their own chart to analyze data and formulate conclusions.

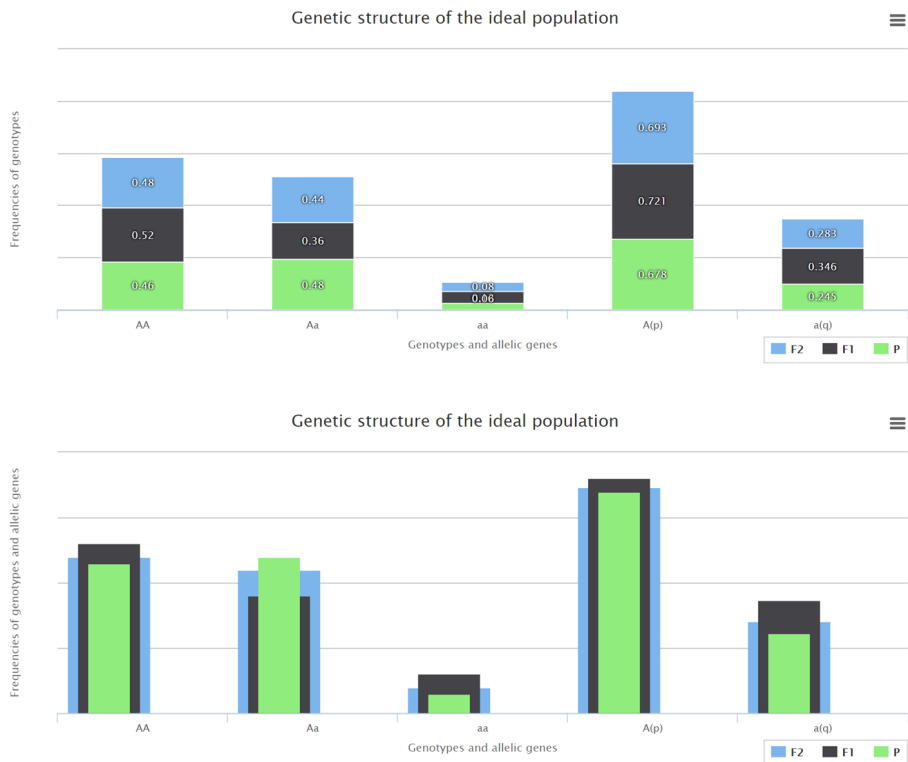

**Fig. 5.** View of diagrams on the web page that were built in automatic mode

Both variants have advantages and disadvantages. In the first variant of the diagram, numeric data of the results of the experiment are included in the corresponding segments of each column. All the data are displayed on the screen, so the student can quite easily compare the numbers.

In the second variant, the segments of each column are located one behind the other, and so that the first, the most narrow segment corresponds to the parent generation and the last, the widest one corresponds to the last child generation. This way of presenting data is liked by students because, not even using numerical data it is visually easy to compare the size (height) of colored bars. Besides, when one aims the cursor at the corresponding field the necessary numerical information appears on the screen.

Analysis of the received data of the model experimentation by the students is carried out on the basis of the analysis of the built:

1. graphics of genetic and genotype structure of the population in generations;

2. one of the diagrams of the genetic structure of the population in generations;
3. graphs and diagrams that overlap.

A variety of graphic options allows to acquaint students with the methods of their statistical processing and presentation.

## 3   Conclusions and Outlook

Modelling of biological processes among population with the means of computer technology is an effective method to develop a series of genetic and evolutionary concepts and results in savings of time resources in the classroom. The use of computer technology as a means of modelling contributes to the formation of the concepts about the possibility of application of elementary statistical methods in biological research, understanding of the nature of statistical laws, in particular the law of large numbers.

In the discussion of the results of the model experiment performed using the developed web pages, the teacher focuses on a small size of the model population (about 50 individuals). For the experiment, one can take a smaller or larger number of individuals, but note that, on the one hand, the smaller the sample size is, the greater the error in the calculations may be. On the other hand, the feature of such studies is that under the conditions of school experimentation with the training model an ideal population for the implementation of practical actions with tangible objects – the models of alleles – it is impossible to comply with the such a condition of validity of the law of Hardy – Weinberg as a large population size. Theoretically it is possible to take this condition into account, if the move away from practical handling of material objects, replacing it with a fully automated process of determining the genetic (genotype and genetic) structure of the population. Students will enter manually data on the number of investigated parental populations and the output frequency of allelic genes in it to the model. We are its research without the use of materialized models. We have begun work in this direction and created the web page http://mybio.education/mod/exp6/en/index.html#. Its use in the teaching of biology does not require simulation with persisted models.

### 3.1   Acknowledgments

The author is grateful to Ariyenchuk Serhiy for advice and technical support in the development and improvement of the functionality of web pages for the modelling of genetic and evolutionary processes in populations.

## References

1. Hammersmith, R.L., Mertens, T.R.: Teaching the Concept of Genetic Drift Using a Simulation. The American Biology Teacher. **52**(8), 497–499 (1990)
2. Mertens, T.R.: Introducing Students to Population Genetics and the Hardy-Weinberg Principle. The American Biology Teacher. **54**(2), 103–108 (1992)

3. Mertens, T.R.: Introducing Students to Population Genetics and the Hardy-Weinberg Principle. In: Moore, R. (ed.) Biology Labs That Work: The Best of How-To-Do-Its, pp. 171–179. National Association of Biology Teachers, Reston (1994)
4. Maret, T.J., Rissing, S.W.: Exploring Genetic Drift & Natural Selection Through a Simulation Activity. The American Biology Teacher. **60**(9), 681–683 (1998)
5. Mukhopadhyay, C.C., Henze, R., Moses, Y.T.: How Real is Race? A Sourcebook on Race, Culture, and Biology (2nd Edition). AltaMira Press, Lanham (2014)
6. Pongsophon, P., Roadrangka, V., Campbell, A.: Counting Buttons: demonstrating the Hardy-Weinberg principle. Science in School. 6, 30–35 (2007)
7. Komarova, O.V.: Modelni eksperymenty pid chas vyvchennia zakonu Khardi – Vainberha (Model experiments during the study of the law of Hardy – Weinberg). Biolohiia i khimiia v suchasnii shkoli (Biology and chemistry in the modern school). 6, 25–31 (2013)
8. Komarova, O.V.: Modelni eksperymenty pid chas vyvchennia zakonu Khardi – Vainberha (Model experiments during the study of the law of Hardy – Weinberg). Biolohiia i khimiia v suchasnii shkoli (Biology and chemistry in the modern school). 4, 19–25 (2013)